\documentclass{iaus}
\usepackage{graphicx}
\usepackage{lscape}

\title[The Arcturus Moving Group] 
{The Arcturus Moving Group: \\ Its Place in the Galaxy}

\author[Mary E. K. Williams, Ken C. Freeman, Amina Helmi and the RAVE collaboration]   
{Mary E. K. Williams $^{1,2}$, Ken C. Freeman$^1$, Amina Helmi$^3$ and the RAVE collaboration}

\affiliation{$^1$ Mt Stromlo Observatory, Cotter Road, Weston Creek, ACT 2611, Australia, \newline 
$^2$ Astrophysikalisches Institut Potsdam, An der Sternwarte 16, D-14482, Potsdam, Germany, \newline $^3$ Kapteyn Institute, P.O. Box 800, 9700 AV Groningen, the Netherlands \\ email: {\tt mary@aip.de};
\\ email: {\tt kcf@mso.anu.edu.au}; \\ email: {\tt ahelmi@astro.rug.nl}}

\pubyear{2008}
\volume{254}  
\pagerange{1--5}
\jname{The Galaxy Disk in Cosmological Context}
\editors{J. Andersen, J. Bland-Hawthorn and B. Nordstr\"om, eds.}
\begin{document}

\maketitle

\begin{abstract}
The Arcturus moving group is a well-populated example of phase-space substructure within the disk of our Galaxy. With its large rotational lag ($V=-100\ \textrm{kms}^{-1}$), metal-poor nature ($\textrm{[Fe/H]}\sim-0.6$) and significant age (10 Gyr) it belongs to the Galaxy's thick disk. Traditionally regarded as the remains of a dissolved open cluster, it has recently been suggested to be a remnant of a satellite accreted by our Galaxy. 

We confirm via further kinematic studies using the \cite[Nordstr\"om et al. (2004)]{Nordstrom2004}, \cite[Schuster et al. (2006)]{Schuster2006} and RAdial Velocity Experiment surveys (RAVE, \cite[Steinmetz et al. 2006]{Steinmetz2006}) the existence of the group, finding it to possibly favour negative $U$ velocities and also possibly a solar-circle phenomenon. We undertook a high-resolution spectroscopic abundance study of Arcturus group members and candidates to investigate the origin of the group. Examining abundance of Fe, Mg, Ca, Ti, Cr, Ni, Zn, Ce, Nd, Sm and Gd for 134 stars we found that the group is chemically similar to disk stars and does not exhibit a clear chemical homogeneity.

The origin of the group still remains unresolved: the chemical results are consistent with a dynamical origin but do not entirely rule out a merger one. Certainly, the Arcturus group provides a challenge to our understanding of the nature and origin of the Galaxy's thick disk.

\end{abstract}

\section{Introduction}

The Arcturus moving group was discovered by Eggen who over the years gathered a list of stars kinematically associated with Arcturus (\cite[Eggen 1971, 1998 and references therein]{Eggen1971, Eggen1998}). Eggen based group membership mainly upon the common $V$ space velocity, the component of stellar motion relative to the Local Standard of Rest (LSR) in the direction of rotation. For the Arcturus moving group the stars lag the LSR with $V=\sim-100\ \textrm{kms}^{-1}$. While Eggen's analysis was somewhat controversial as he adjusted the stellar parallaxes to return a tight $V$-velocity relation, \textit{a posteriori} justification was obtained by a tight colour-magnitude relation along an isochrone with age $\tau\gtrsim10$ Gyr and metallicity $\textrm{[Fe/H]}\sim-0.6$ (\cite[Eggen 1996]{Eggen1996}). The age, metallicity and space velocity identify the group as part of the Galaxy's thick disk. 

Eggen introduced the Arcturus group as a dissolved open cluster. In this scenario a single star-forming event creates a cluster of stars which over time dissolves. Eventually, the most distinguishing characteristic of the stars' common origin are their similar space motions. However, recently it was suggested by \cite[Navarro et al. (2004)]{Navarro2004} that the Arcturus moving group could be an example of the debris of an accreted satellite in the disk of the Galaxy. If this indeed is the case, this would be of significance in the debate about the origin of the thick disk; if the Arcturus group is an example of accretion debris this would lend weight to the argument that the thick disk consists mainly of such debris. 

The motivation behind this study was therefore to further understand the formation of the Arcturus group in the context of thick disk formation; is the group a remnant of a star-formation event in the disk or accretion debris? Our primary approach was to perform a high-resolution spectroscopic abundance program to search for tracers of the origin of the group in the chemistry of the stars. Following this study, we turned to kinematic studies to confirm the group's existence and further define its kinematic properties. In the following we briefly report on the intriguing results of this study, which lead to more questions than answers about the origin of the Arcturus moving group. A more detailed analysis will be presented shortly in \cite[Williams et al. (2008)]{Williams2008}. 

\section{Abundance study} \label{sec:Abundances}
\subsection{Candidate Selection} \label{subsection:Candidate}
In light of the possible extra-Galactic origin of the Arcturus group there was the possibility that members could have been missed by Eggen as they might not satisfy his strict selection criterion. So in addition to Eggen's list of Arcturus stars we selected group candidates from the RAVE (\cite[Steinmetz 2006]{Steinmetz2006}), \cite[Nordstr\"om et al. (2004), Beers et al. (2000)]{Nordstrom2004, Beers2000} and \cite[Norris (1986)]{Norris1986} studies. To develop selection criteria inclusive of both formation scenarios we performed \textit{N}-body simulations of the formation of the Arcturus group via dissolution of a progenitor in a static Galactic potential with a range of progenitor masses. The simulations follow those presented in \cite[Helmi (2006)]{Helmi2006} and
 \cite[Navarro (2004)]{Navarro2004} where a satellite or cluster with an orbit similar to that of
 the Arcturus is disrupted in the potential of the Galaxy. The largest of these, the dissolved dwarf spheroidal galaxy, was used to develop the ``banana" criterion selecting stars that are near or at the apocentre of their orbits falling within a banana-shaped region in the $UV$ plane around $V=-100\ \textrm{kms}^{-1}$. Also, we restricted our study to stars on disk-like orbits with $|W|< 100\ \textrm{kms}^{-1}$ and with small velocity errors. Some stars from the thick and thin disks were included for comparative purposes. A total of 134 stars from 190 observations were analysed in our study. 

\subsection{Methodology} \label{subsection:methodology}
Elemental abundances were derived for our candidate stars by performing a Local Thermodynamic Equilibrium (LTE) analysis with the MOOG code (\cite[Sneden 1973]{Sneden1973}) on our high resolution, high signal-to-noise UCLES data obtained in three observing periods at the AAT from August 2003 - November 2006. The first of these runs was obtained by Gayandhi de Silva and kindly granted to us for analysis. To measure equivalent widths we used the new DAOSPEC program which automatically fits Gaussian profiles to lines in a spectrum (\cite[Pancino \& Stetson 2008]{Pancino2008}). However, it was necessary to alter the DAOSPEC code with regards to continuum fitting of the spectra; most of our data is in the blue and so very crowded and it was found that DAOSPEC generally set the continuum too low. The DAOSEC continuum fitting was therefore disabled and hand fits performed. The line list was compiled from the literature utilising laboratory $\log gf$ values where available. Abundances were derived for Fe, Mg, Ca, Ti, Cr, Ni, Zn, Ce, Nd, Sm and Gd for our 134 stars using primarily spectra in the blue. For a subset of these stars we also have red data from which we obtained abundances for Na, Mg, Al and Si. Here we present only those elements or lines for which hyperfine and isotopic splitting need not be considered. 

Stellar parameters were calculated using `physical' and `spectroscopic' approaches. The former involved calculating $T_{\scriptstyle\textrm{eff}}$ and $\log g$ from photometric and astrometric data, while the latter utilises forcing excitation and ionisation balance to derive $T_{\scriptstyle\textrm{eff}}$ and $\log g$ respectively. In both cases the microturbulence is found by requiring that there is no dependence of abundance on line strength. In this short report we only include the spectroscopic results as they yielded better agreement with abundance for stars in common with the studies of \cite[Reddy et al. (2003, 2006)]{Reddy2003, Reddy2006} and \cite[Bensby et al. (2003, 2005)]{Bensby2003, Bensby2005}. The difference in abundance between stars in common with those studies and our own are $< \textrm{[Fe/H]}_{this\ study}-\textrm{[Fe/H]}_{Reddy}>\ =0.09$ with a standard deviation of $\sigma=0.05$, and for $< \textrm{[Fe/H]}_{this\ study}-\textrm{[Fe/H]}_{Bensby}>\ =0.05$ with a standard deviation of $\sigma=0.07$. Other elements have comparable errors. The agreement is good considering the different techniques, line lists and calibrations employed.

\subsection{Results} \label{subsection:abundance_results}
\begin{figure}[t!]
\begin{center}
 \includegraphics[width=\textwidth]{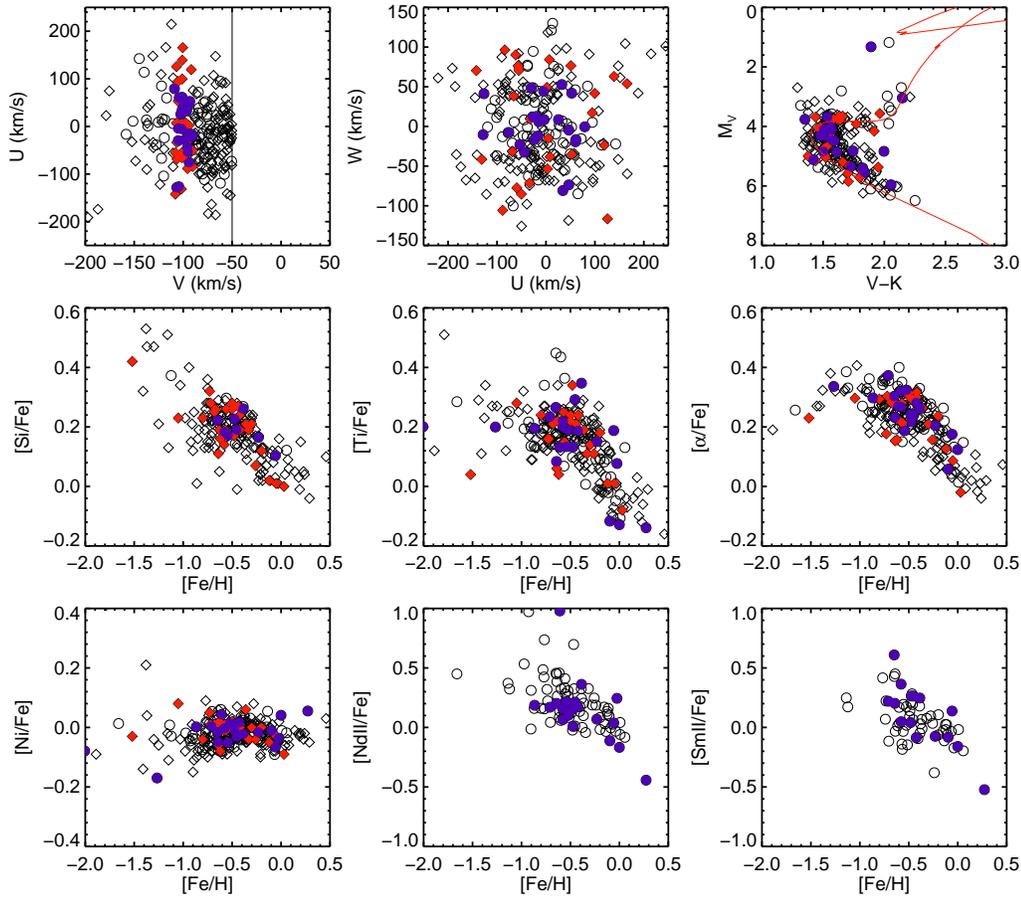} 
 \caption{The $VU$, $UW$ planes, CMD and abundance ratio plots for the Arcturus group candidates (blue circles and red diamonds) selected from stars with $V<-50\ \textrm{kms}^{-1}$. Stars from this study are designated by circles, while the amalgamated Reddy and   samples are diamonds. The overlaid Padova isochrone in the CMD has $12.5\ \textrm{Gyr}$, $Z=0.006$. }
 \label{fig1}
\end{center}
\end{figure}

In Section \ref{subsection:schuster_GC} we will see that the Arcturus group is an over-density in a very narrow velocity range around $V=-100\ \textrm{kms}^{-1}$, with a standard deviation of only $\sigma_V=3\ \textrm{kms}^{-1}$. We therefore revert here to a selection mimicking Eggen's initial criteria, selecting as Arcturus candidate stars those that are $\pm10\ \textrm{kms}^{-1}$ from this mean V velocity. Also, to compare the group against the background thick disk stars we include stars with $V< -50\ \textrm{kms}^{-1}$. Figure \ref{fig1} shows selected abundance results for the Arcturus candidates superimposed on the background stars. The results from the \cite[Reddy et al. (2003, 2006)]{Reddy2003, Reddy2006} and \cite[Bensby et al. (2003, 2005)]{Bensby2003, Bensby2005} studies are included to further emphasise the general trends of field stars and to increase the number of candidate stars. The largest of the systematic differences between these studies and our results were accounted for by using the zero-point offsets derived from common stars to shift their results to our scale.

We see clearly in these plots that within the limit of our abundance errors there is no distinguishing features in abundance between the Arcturus group stars and those of the surrounding disk. While Nd suggests a possible clustering of abundance, this is not corroborated by the $\alpha$- or other elements. Also, the apparently tight CMD relation of Eggen now seems like a sampling from a old population of similar age, i.e., the thick disk, as both the Arcturus candidates and the background stars lie
reasonably tightly along the same isochrone. Employing different selection criteria for the group, such as our banana selection criteria mentioned above or selecting Eggen's original Arcturus group candidates yields the same result: we are unable to find clear distinguishing features in abundance for the Arcturus group when compared to background disk stars.

These results beg the question: \textit{does the Arcturus moving group exist at all as an over-density in phase space}? So before drawing any further conclusions we turn to kinematic studies to confirm the group's existence and define it kinematically.

\section{Kinematic study} \label{subsection:schuster_GC}
While the Geneva-Copenhagen catalog provided a 
significant number of Arcturus candidates, it is not the ideal source 
because it has relatively few stars at the high-velocity of the group. However, the recent study of solar-neighbourhood metal-poor stars by \cite[Schuster et al. (2006)]{Schuster2006} provides a wealth of stars in the thick disk. This study was not available to us at the time of choosing candidates for our abundance study, however we can now use it to investigate the phase-space structure in the vicinity of Arcturus. Note that there are kinematical selection biases in the Schuster data set towards high-proper motion stars which means that stars with a $V$ velocity nearer to the sun ($V\sim0$) are under-represented. However, the region around the Arcturus group's velocity is not affected by these biases. 

\begin{figure}[ht!]
\begin{center}
 \includegraphics[width=\textwidth]{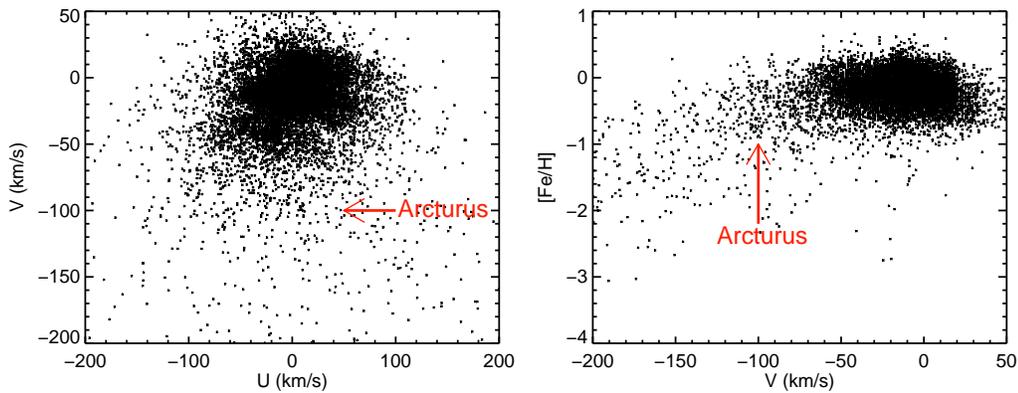} 
 \caption{The Arcturus group as seen in the combined \cite[Nordstr\"om et al. (2004)]{Nordstrom2004} and \cite[Schuster et al. (2006)]{Schuster2006} data set.}
 \label{fig2}
\end{center}
\end{figure}

Figure \ref{fig2} displays plots of kinematic and metallicity plots for the combined Nordstr\"om and Schuster data sets where stars in common appear only once. In this diagram we see an over-density at $V\sim-100\ \textrm{kms}^{-1}$ extending from $0<\textrm{[Fe/H]}<-1$. This feature is particularly prominent at lower metallicities. \cite[Schuster et al. (2006)]{Schuster2006} interpreted this over-density as the thick disk being split into two components, with the split indicative of the thick disk consisting of merger debris. They link it with the \cite[Gilmore, Wyse \& Norris (2002)]{Gilmore2002} feature but do not associate it to the Arcturus group. We see however that the `second component' of the thick disk that they identify is clearly at the Arcturus group velocity. Combining their data with the Nordstr\"om sample shows that the over-density is perhaps not as localised in [Fe/H] as their results indicated. 

We see also that there is a suggestion of a bias towards negative $U$, which is coincident with the favoured $U$ of the Hercules group. From fits to the generalised histogram in the $V$-velocity we find that the group is centred around $\mu_V=-102\ \textrm{kms}^{-1}$ with a very narrow velocity dispersion of $\sigma_V=2-3\ \textrm{kms}^{-1}$. 

We have also investigated the Arcturus group in RAVE data, utilising the internal release of October 2007 which contains 191,170 radial velocity measurements as well as stellar parameters for a significant number of stars (see \cite[Zwitter et al. 2008]{Zwitter2008}). The stellar parameters enabled us to select those stars associated with the helium-burning red clump. The $K$-band magnitude of the clump, while being relatively unaffected by extinction, has also been shown observationally to be relatively independent of metallicity and age (\cite[Pietrzy{\'n}ski, Gieren \& Udalski 2003, Alves 2000]{Pietrzynski et al. 2003, Alves et al. 2000}). Thus, combining RAVE data with photometric and astrometric data we were able to derive reliable distances and $UVW$ velocites for a sample of 16,000 red clump giants. In this data set we have found that the Arcturus group seems to be more pronounced in the solar circle, i.e., for those with a Galactocentric radius similar to the sun (for full details see \cite[Williams et al. 2008]{Williams2008}). If this is indeed the case, such behaviour could be possibly be explained by a dynamical origin of the group. 

\section{Discussion} \label{Discussion}

\cite[Venn et al. (2004)]{Venn2004} demonstrated that current-day satellite dSph galaxies of the Milky Way are $\alpha$-poor when compared to the Galactic stars of similar [Fe/H], indicating a comparatively slow star formation history (\cite[Matteucci 2003]{Matteucci2003}). So the similarity between the Arcturus group members and surrounding disk weighs against the accretion debris origin of the group \textit{if we expect the Arcturus progenitor to be similar to current day dSphs}. However, from \cite[Tamura, Hirashita \& Takeuchi (2001)]{Tamura2001} we see that a satellite progenitor of Arcturus would have had to have a substantial mass (similar to the LMC) to be enriched to [Fe/H]=-0.6. Therefore, it is possible that a such a large dwarf galaxy had a much higher rate of star formation than the remaining satellites of the Milky Way. Also, the simulations of \cite[Abadi (2003)]{Abadi2003} of the formation of a Milky Way-like galaxy in the $\Lambda$CDM cosmogony gave 60\% of the thick disk as tidal debris. So our results showing the Arcturus group as being similar to disk stars does not entirely exclude the possibility that it, and the surrounding thick disk, are tidal debris. However, it would seem remarkable that a disparate group of satellites could conspire to have such similar abundances; a clear picture of the chemical evolution of these satellites to produce the observed abundance patterns has yet to be drawn. 

If we instead suppose an \textit{in situ} formation of the thick disk, such as that proposed by \cite[Brook et al. (2005)]{Brook2005} from gas-rich mergers, the chemical similarity of the Arcturus group to the surrounding disk favours \textit{in situ} formation of the group as well. However, from our simulations we find that even the largest open clusters ($7 \times 10^4\ \rm{M_\odot}$) could not produce a discernible over-density in the solar neighbourhood after $10\ \textrm{Gyr}$ of evolution (see \cite[Williams et al. (2008)]{Williams2008}). Furthermore, stars in the Arcturus velocity range do not exhibit the chemical homogeneity expected from a dissolved star forming event (e.g. see \cite[de Silva et al. (2007)]{deSilva2007}). Instead we see that our results are similar to \cite[Bensby et al. (2007)]{Bensby2007} for the Hercules moving group where they found that the group is chemically indistinct from the disk. This supports the dynamic origin of the thick/thin disk Hercules group, which is thought to arise due to the Outer Lindblad Resonance with the Galactic bar (\cite[Dehnen 2000]{Dehnen2000}, \cite[Fux 2001]{Fux2001}). We therefore wonder, could this also be the case for the Arcturus group? This scenario could be supported by the Arcturus group seemingly mimicking Hercules in being asymmetrical in $U$ and exhibiting a dependence on Galactocentric radius. Also, the $V$ velocity of Arcturus is coincident with that of the 6:1 OLR of the bar at the solar position. However, how such a resonance could produce an over-density, affecting stars that spend so much time out the Galaxy's plane is not currently understood. 

We are continuing our investigations of the Arcturus group. In our upcoming paper we will present our abundance results in full, as well as including those for extra elements. We will give also present the full kinematic results, comparing them to our simulations of various progenitor scenarios and explore the resonance possibility further. The Arcturus group's origin is an enigma still yet to be solved, but the clues are accumulating.


\begin{thebibliography}{}

\bibitem[Abadi et al. (2003)]{Abadi2003}
{Abadi, M.G., Navarro, J.F., Steinmetz, M. \& Eke, V.R.}
2003, \textit{ApJ}, 597, 21
 
\bibitem[Alves (2000)]{Alves2000}
{Alves, D.R.}
2000, \textit{ApJ}, 539, 732
 
\bibitem[Beers et al. (2000)]{Beers2000}
{Beers, T.C.,  Chiba, M., Yoshii, Y., Platais, I., Hanson, R.B., Fuchs, B. \& Rossi, S.}
2000, \textit{AJ}, 119, 2866

\bibitem[Bensby, Feltzing \& Lundstr\"om (2003)]{Bensby2003}
{Bensby, T., Feltzing, S. \& Lundstr\"om, I.}
2003, \textit{A\&A}, 410, 527

\bibitem[Bensby et al. (2005)]{Bensby2005}
{Bensby, T., Feltzing, S., Lundstr\"om, I. \& Ilyin, I.}
2005, \textit{A\&A}, 433, 185

\bibitem[Bensby et al. (2007)]{Bensby2007}
{Bensby, T., Oey, M.S., Feltzing, S. \& Gustafsson, B.}
2007, \textit{ApJ}, 655, L89

\bibitem[Brook et al. (2005)]{Brook2005}
{Brook, C.B., Gibson, B.K., Martel, H. \& Kawata, D.}
2005, \textit{ApJ}, 630, 298

\bibitem[de Silva et al. (2006)]{deSilva2006}
{De Silva, G.M., Sneden, C., Paulson, D.B., Asplund, M., Bland-Hawthorn, J., Bessell, M.S. \& Freeman, K.C.}
2006, \textit{AJ},131, 455
    
\bibitem[Dehnen (2000)]{Dehnen2000}
{Dehnen, W.}
2000, \textit{AJ}, 119, 800

\bibitem[Eggen (1971)]{Eggen1971}
{Eggen, O.}
1971, \textit{PASP}, 83, 271

\bibitem[Eggen (1996)]{Eggen1996}
{Eggen, O.}
1996, \textit{AJ}, 112, 1595

\bibitem[Eggen (1998)]{Eggen1998}
{Eggen, O.} 1998,
\textit{AJ}, 115, 2397

\bibitem[Fux (2001)]{Fux2001}
{Fux, R.}
2001, \textit{A\&A}, 373, 511

\bibitem[Gilmore, Wyse \& Norris (2002)]{Gilmore2002}
{Gilmore, G., Wyse, R.F.G. \& Norris, J.E.}
2002, \textit{ApJ}, 574, L39

\bibitem[Helmi et al. (2006)]{Helmi2006}
{Helmi, A., Navarro, J.F., Nordstr{\"o}m, B., Holmberg, J., Abadi,
  M.G., and Steinmetz, M.}
2006, \textit{MNRAS}, 365, 1309

\bibitem[Matteucci (2003)]{Matteucci2003}
{Matteucci, F.}
2003, \textit{Ap\&SS}, 284, 539

\bibitem[Navarro, Helmi \& Freeman (2004)]{Navarro2004}
{Navarro, J.F., Helmi, A. \& Freeman, K.C.}
2004, \textit{ApJ}, 601, L43

\bibitem[Nordstr\"om et al. (2004)]{Nordstrom2004}
{Nordstr{\"o}m, B., Mayor, M., Andersen, J., Holmberg, J., Pont, F., J{\o}rgensen, B.R., Olsen, E.H., Udry, S. \& Mowlavi, N.}
2004, \textit{A\&AS}, 418, 989

\bibitem[Norris (1986)]{Norris1986}
{Norris, J.}
1986, \textit{ApJS}, 61, 667

\bibitem[Pacino \& Stetson (2008)]{Pacino2008}
{Pacino \& Stetson}
2008, in preparation

\bibitem[Pietrzy{\'n}ski, Gieren \& Udalski (2003)]{Pietrzynski2003}
{Pietrzy{\'n}ski, G., Gieren, W., and Udalski, A.}
2003, \textit{AJ}, 125, 2494.

\bibitem[Reddy, Lambert \& Allende Prieto (2006)]{Reddy2006}
{Reddy, B.E., Lambert, D.L. \& Allende Prieto, C.}
2006, \textit{MNRAS}, 367, 1329

\bibitem[Reddy et al. (2003)]{Reddy2003}
{Reddy, B.E., Tomkin, J., Lambert, D.L. \& Allende Prieto, C.}
2003, \textit{MNRAS}, 340, 304

\bibitem[Schuster et al. (2006)]{Schuster2006}
{Schuster, W.J., Moitinho, A., M\'arquez, A., Parrao, L. \& Covarrubias, E.}
2006, \textit{A\&A}, 445, 939

\bibitem[Sneden (1973)]{Sneden1973}
{Sneden, C.}
1973, PhD thesis, University of Texas, Austin

\bibitem[Steinmetz et al. (2006)]{Steinmetz2006}
{Steinmetz, M. et al.}
2006, \textit{AJ}, 132, 1645

\bibitem[Tamura, Hirashita \& Takeuchi (2001)]{Tamura2001}
{Tamura, N., Hirashita, H. \& {Takeuchi}, T.T.}
2001, \textit{ApJ},  552, 113

\bibitem[Venn et al. (2004)]{Venn2004}
{Venn, K.A., Irwin, M., Shetrone, M.D., Tout, C.A., Hill, V. \& Tolstoy, E.}
2004, \textit{AJ}, 128, 1177

\bibitem[Williams et al. (2008)]{Williams2008}
{Williams, M. E. K et al.}
2008, in preparation


\bibitem[Zwitter et al. (2008)]{Zwitter2008}
{Zwitter, T. et al.}
2008, \textit{AJ}, 136, 421


\end{thebibliography}
\end{document}